\documentclass[conference]{IEEEtran}
\IEEEoverridecommandlockouts
\usepackage{cite}
\usepackage{amsmath,amssymb,amsfonts}
\usepackage{algorithmic}
\usepackage{graphicx}
\usepackage{hyperref}
\usepackage{textcomp}
\usepackage{xcolor}
\def\BibTeX{{\rm B\kern-.05em{\sc i\kern-.025em b}\kern-.08em
    T\kern-.1667em\lower.7ex\hbox{E}\kern-.125emX}}

\newcommand{\etal}{\textit{et al.}}

\newcommand\blfootnote[1]{%
  \begingroup
  \renewcommand\thefootnote{}\footnote{#1}%
  \addtocounter{footnote}{-1}%
  \endgroup
}

\begin{document}
\title{Ultra-Low Power Keyword Spotting at the Edge}

\author{\IEEEauthorblockN{Mehmet Gorkem Ulkar}
\IEEEauthorblockA{\textit{Analog Devices Inc.}\\
Istanbul, Turkey \\
gorkem.ulkar@analog.com}
\and
\IEEEauthorblockN{Osman Erman Okman}
\IEEEauthorblockA{\textit{Analog Devices Inc.}\\
Istanbul, Turkey \\
erman.okman@analog.com}
}

\maketitle

\begin{abstract}
Keyword spotting (KWS) has become an indispensable part of many intelligent devices surrounding us, as audio is one of the most efficient ways of interacting with these devices. The accuracy and performance of KWS solutions have been the main focus of the researchers, and thanks to deep learning, substantial progress has been made in this domain. However, as the use of KWS spreads into IoT devices, energy efficiency becomes a very critical requirement besides the performance. We believe KWS solutions that would seek power optimization both in the hardware and the neural network (NN) model architecture are advantageous over many solutions in the literature where mostly the architecture side of the problem is considered. In this work, we designed an optimized KWS CNN model by considering end-to-end energy efficiency for the deployment at MAX78000, an ultra-low-power CNN accelerator. With the combined hardware and model optimization approach, we achieve 96.3\% accuracy for 12 classes while only consuming 251 uJ per inference. We compare our results with other small-footprint neural network-based KWS solutions in the literature. Additionally, we share the energy consumption of our model in power-optimized ARM Cortex-M4F to depict the effectiveness of the chosen hardware for the sake of clarity. \blfootnote{Preprint.}

% For the energy efficiency in hardware front, latest deep neural network (DNN) and convolutional neural network (CNN) accelerators provide massive energy and latency efficiencies. Yet, these accelerators are resource constrained; both the supported architectural components and the deployable model sizes are limited. Furthermore, some applications include preprocessing or post processing steps where these accelerators cannot be used and less energy efficient general purpose microcontrollers (MCUs) need to be utilized for these actions. This decreases the overall end-to-end energy efficiency.

\end{abstract}

\begin{IEEEkeywords}
speech commands, keyword spotting, wake word detection, convolutional neural networks, once-for-all, neural architecture, mfcc
\end{IEEEkeywords}

\section{Introduction}
Speech-based interfaces have been commonly used for smart devices like intelligent loudspeakers or mobile assistants, especially with the recent improvements in deep neural network (DNN) technologies. Large vocabulary speech recognition systems run big networks that require a lot of resources and high processing power; therefore, they are generally built as cloud-based solutions. On-device keyword spotting (KWS) is used in such systems to activate or support a limited number of interactions (i.e., commands) to reduce standby energy and latency, preserve user privacy, and avoid data leakage during data transmission. 

Due to the low-latency requirement of KWS systems, many researchers focused on solutions with a small footprint in terms of computation and parameters. The authors in \cite{sainath2015convolutional} proposed using a CNN with limited size instead of a DNN for the KWS application. In \cite{tang2018deepresidual}, the authors proposed the use of residual layers and dilations in CNN to increase the accuracy while satisfying the small footprint requirement. Graph convolutional networks are proposed in \cite{chen2019smallfootprint} to capture non-local relations in KWS. Zhang \etal proposed a depthwise-separable CNN for MCU deployment in \cite{zhang2018hello}. All of these mentioned methods use Mel-Frequency Cepstrum Coefficients (MFCC) to preprocess the raw audio first and then run their models on these features. While they are competing to get more compact and accurate models, the computation and power consumed in the MFCC 
remain stable. The authors in \cite{Tsetlin} proposes using Tsetlin Machines instead of NN to run on MFCC features, and they show that MFCC consumes most of the total energy of their end-to-end KWS system. This illustrates the significance of eliminating MFCC or other similar preprocessing from the KWS pipeline. Authors in \cite{mittermaier2020kwswithsincconv} proposes using Sinc-convolutions in the feature extraction layer of their CNN network instead of MFCC. However, Sinc-convolutions are not supported by most of the known NN accelerators, if not none. Therefore this block needs to be implemented in the MCU if edge deployment is targeted.

% The main contribution of this work is threefold:
% \begin{itemize}
% \item A network architecture is proposed for KWS applications that runs on time domain signals to reduce the required computation resources while preserving the accuracy.
% \item An ultra-low power CNN accelerator, MAX78000, is utilized to implement the solution to present the overall energy reduction.
% \item A OFA based NAS approach is implemented and used to maximize the performance of the proposed network architecture while considering the MAX78000 hardware limitations.
% \end{itemize}

\begin{figure}[h]
\centering
\includegraphics[width=0.6\linewidth]{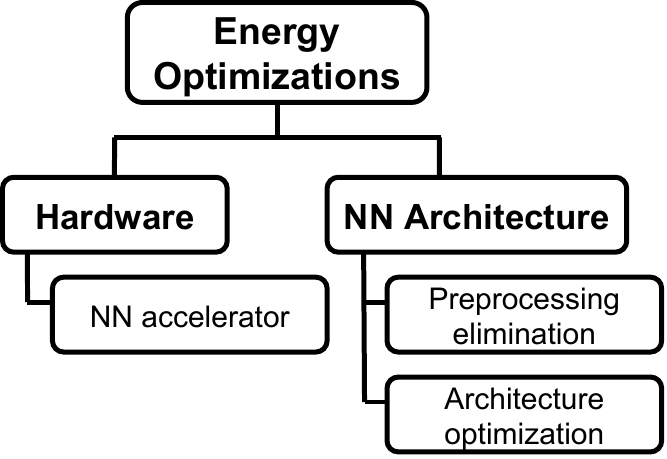}
\caption{Energy optimization methods}
\label{Fig1}
\end{figure}

Contrary to most literature where the number of operations and parameter number are the main concerns, energy is the most vital consideration for any battery-powered edge solution. With the introduction of NN accelerators, the metrics like multiply-accumulate operations (MAC) do not substitute for energy usage as the energy usage of those operations differs vastly depending on whether those operations can be done in an ultra-efficient NN accelerator or not. The main focus of this paper is to propose a highly accurate and ultra-energy-efficient KWS solution. For the lowest possible end-to-end energy usage, both the hardware and model architecture need to be optimized as shown in Fig. \ref{Fig1}. The utilization of specialized low-power hardware for neural network inference is the general solution if custom hardware is not to be designed for the given application. However, most of the time, neural networks are not the single components of the keyword spotting systems. Preprocessing the audio signal or postprocessing the model outputs could be required besides the neural network inference operation. Removing those parts that cannot be run by an efficient accelerator and delegating their tasks to neural network blocks would be a viable method for increasing overall energy efficiency. Another room for improvement in the architecture side is to obtain the best architecture in terms of performance and energy for the given limitations of the deployment hardware.

In this study, we propose a CNN-based neural network architecture for KWS problems, which reduces the required energy consumption by removing the commonly used Mel-Spectrum transformation and utilizing an optimized neural network accelerator, MAX78000. We also employ a neural architecture search (NAS) approach and modify due to the limitations of the specialized hardware to optimize the accuracy of the final model. To summarize, our contributions can be listed as follows:
\begin{itemize}
\item In the hardware side, an ultra-low-power CNN accelerator, MAX78000, is utilized to implement the solution to present the overall energy reduction.
\item In the NN architecture front, a network architecture is proposed for KWS applications that run on time-domain signals to reduce the required computation resources while preserving the accuracy.
\item Once-For-All (OFA) based NAS approach is implemented and used to maximize the performance of the proposed network architecture while considering the MAX78000 hardware limitations.
\item Proposed solution is compared with other keyword spotting methods in the literature regarding performance and energy consumption.
\end{itemize}

% \begin{figure}[h]
% \centering
% \begin{adjustbox}
% \includegraphics[width=0.6\linewidth]{figures/HwSw_Optimizations.pdf}
% \caption{Caption 1}
% \label{Fig1}
% \end{adjustbox}
% \end{figure}   

\section{Neural Network Architecture}

\begin{figure}[ht]
\centering
\includegraphics[trim=0 0 0 0,clip, width=1.01\linewidth]{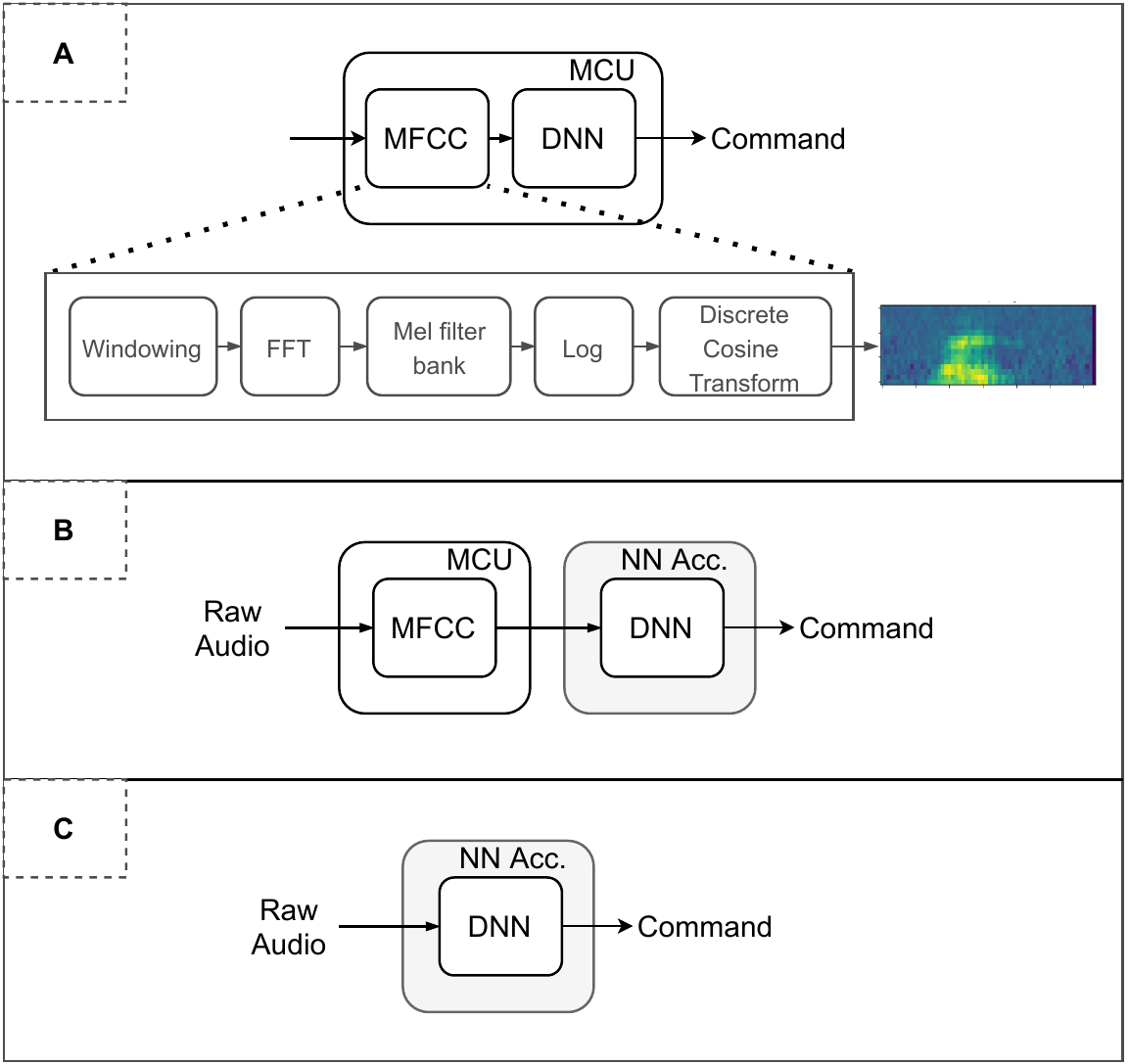}
\caption{KWS approaches}
\label{fig:approaches}
\end{figure}

\subsection{CNN Model with Raw Audio Input}
In many of the keyword spotting applications, MFCC is commonly utilized features \cite{sainath2015convolutional}, \cite{tang2018deepresidual}, \cite{zhang2018hello} which represents the speech signal with a set of known and relevant components. MFCC are achieved by decomposition of the audio signal using a filter bank. It provides the Discrete Cosine Transform (DCT) of a real logarithm of the short-term energy on the Mel frequency scale. More specifically, the computation pipeline of MFCC includes windowing the speech signal into frames, performing Fast Fourier Transform (FFT) to find the power spectrum of each frame, filter bank processing using the Mel scale, and finally DCT on the log scale of the power spectrum. Therefore, the computational load of this preprocessing step is generally much higher than the inference of the trained networks. 

If the models that require MFCC processing are deployed to NN accelerators, MFCC has to be implemented in a general-purpose MCU outside of the accelerator, as shown in Fig. \ref{fig:approaches} B. However, this reduces the overall efficiency as an important portion of computation does not take place in the ultra-low-power NN accelerator. There are also some efforts in the literature to reduce the power consumption of MFCC processing, but it still requires more than two times more power than the neural network inference \cite{Shan2020}.

\begin{figure*}[h]
  \centering \includegraphics[width=0.75\linewidth]{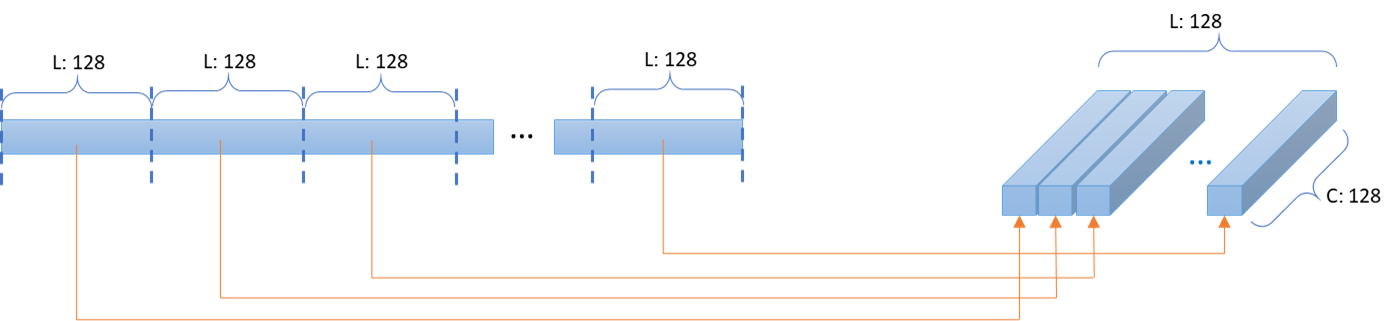}
  \caption{Proposed data folding strategy. 16384x1 signal is reshaped to 128x128x1.}
  \label{fig:datafolding}
\end{figure*}

In this study, we propose to minimize the preprocessing step by elimination of the transformation to generate MFCC to reduce energy consumption effectively, as shown in Fig. \ref{fig:approaches} C. The raw audio signals are fed to the neural network after reshaping the signal, where the short-term and long-term variations can be observed.
We used 1.024s length utterances sampled at 16kHz, and each of them are reshaped to 128x128x1 by folding the raw audio signal to the signal channels as shown in Fig. \ref{fig:datafolding}. Note that this strategy has no processing cost for our NN accelerator as it can be achieved by changing the order of the memory locations where the streaming input data is copied.

The advantage of this data structure is using 1D convolutional operators it is possible to extract features that represent both short time and long time relations. Therefore, in this study, we designed a base network seen in Fig. \ref{fig:nas_network_arch} which is composed of six 1-D convolutional layer units separated by pooling layers and a linear layer at the end. The units are composed of two or three convolutional layers is defined as 128 channel 5x5 kernels, and will then be modified by the NAS as described in the following section. 

\subsection{Neural Architecture Search} \label{section:ofa}
\begin{figure*}[h]
  \centering \includegraphics[width=0.9\linewidth]{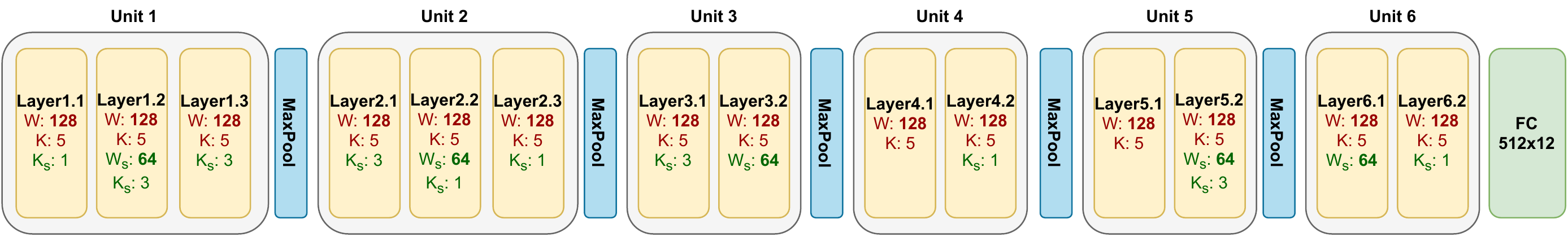}
  \caption{KWS super-network for OFA}
  \label{fig:nas_network_arch}
\end{figure*}

Neural architecture search (NAS) is an increasingly popular research topic in the literature \cite{NAS_Survey}. By automating the exploration process, neural architecture search (NAS) methods aim to find the best neural architecture for a given set of requirements. The models that are found by contemporary NAS methods are difficult to obtain using manual design in terms of complexity and achieved accuracy. In this paper, we employed a weight-sharing-based NAS technique, Once-For-All (OFA) \cite{cai2020once}. The authors in \cite{cai2020once} propose training a super-network and deploying parts of it to diverse hardware without any need for retraining. This super-network is called “Once-for-All” network, and it requires a training process where all sub-networks are trained sufficiently to be deployed directly. Since training all sub-networks is computationally prohibitive, the authors came up with an idea of progressive shrinking. In this training method, the training starts with training the super-network with maximum kernel size, depth and width first and smaller sub-networks that share parameters with super-network become available for sampling progressively. By doing so, the training of interfering sub-networks at the same time is avoided. If the search space consists, different kernel, depth, and width values, they are added to sampling space sequentially. Once the whole training step is completed, the evolutionary search is performed to find the best sub-network for a given set of requirements. The constraints of targeted NN accelerator like memory are considered in this search step. In the Fig. \ref{fig:nas_network_arch}, the designed super-network for the KWS is shown. This super-network contains convolutional layers with maximum width and kernel sizes, and the depth of the network is set to maximum. Specifically, the designed super-network contains 14 identical Conv1D layers with 128 channels and 5 as kernel lengths. Upon OFA training and evolutionary search, the sub-network selected automatically can be shown in the same Fig. \ref{fig:nas_network_arch} with parameter overrides in green font. In case any layer of the automatically selected network has the same parameter value with the super network, it is not indicated with green, so the red default values are used. It is noteworthy that the sub-network is much smaller than the super-network where its parameters are shared with. Another observation is that in this case, OFA did not prefer to eliminate any layer from the super network, although it could do so. Rather than decreasing the depth of the model, it chooses to slim the layers or shorten the kernels.  

According to the OFA scheme, the model is ready to deploy after training and search steps. However, the model needs to be quantized if it is to be deployed to an NN accelerator or any MCU. For this purpose, we employed quantization-aware training (QAT) rather than post-quantization as QAT yields networks with higher performances.

All the codes that have been developed for this study are open-sourced and can be found at \cite{ai8xtraining_repo}.

% \textcolor{blue} {
% \subsection{Quantization Aware Training}
% Should we discuss how QAT works here?}

% \begin{figure*}[h]
%   \centering \includegraphics[width=0.9\linewidth]{figures/kws12_chosen_subnet.pdf}
%   \caption{KWS sub-network found by OFA}
%   \label{fig:nas_chosen_network_arch}
% \end{figure*}

% \begin{figure}[h]
% \centering
% \begin{adjustbox}
% \includegraphics[width=\linewidth]{figures/mofa.drawio_kws1d.pdf}
% \caption{Caption 1}
% \label{Fig1}
% \end{adjustbox}
% \end{figure}   

\section{Experiments \& Results}
\subsection{Dataset}
The proposed approach is evaluated using the Google Speech Commands Dataset v2 \cite{warden2018speech}, which contains 65000 1-second long utterances of 35 short words by thousands of different people, as well as some background noises. Based on the recent researches for KWS applications \cite{tang2017honk, tang2018deepresidual, zhang2018hello, Mo2020}, the dataset is relabeled to form 12 classes: "Yes", "No", "Up", "Down", "Left", "Right", "On", "Off", "Stop", "Go" along with the "Silence" which contains no speech utterances and the "Unknown" which contains the data from the remaining 25 keywords in the original dataset. The dataset is split into training, validation, and test set in the ratio of 80:10:10 while the audio clips from the same person are kept in the same set.

An augmentation procedure is also followed to generate the training data to increase the robustness and the performance of the neural network model. This procedure is composed of signal stretching, utterance shifting, and noise addition, respectively. As a first step, the utterance is stretched with a random factor selected from $\mathcal{U}$(0.8, 1.3). A random time shift, $t$ $ms$, is then applied to the audio signal, where $t$$\backsim\$(\mathcal{U})$(-0.1, 0.1). Lastly, a zero-mean white Gaussian noise is added to the audio signal with variance \(\sigma^2\), which is a uniformly distributed random variable between 0 and 1.

During training, every sample in the dataset is used, so there are more samples from the "Unknown" class than the other class samples. In order to prevent the model from getting biased, the class weight of the "Unknown" samples is reduced proportionally to the amount of the other class samples.

\subsection{Model Training}
\begin{figure}[h]
  \centering \includegraphics[width=0.9\linewidth]{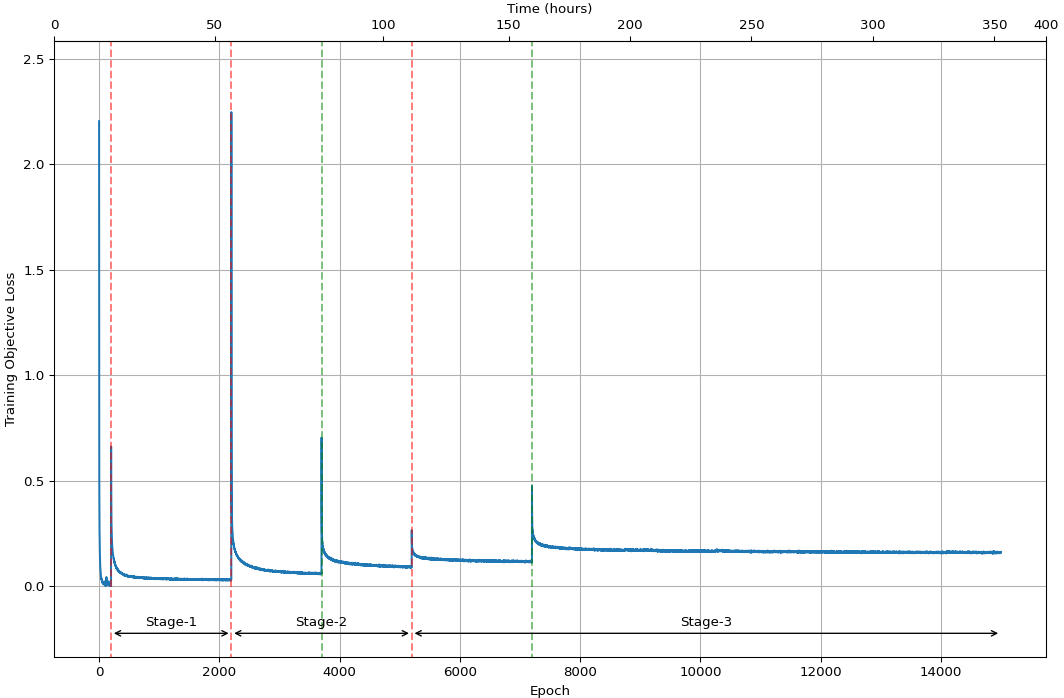}
  \caption{Objective Loss vs. Number of Epochs for training OFA supernet}
  \label{fig:nas_training_plot}
\end{figure}

In this study, model training includes two steps. As the first step, the supernet given in Fig. \ref{fig:nas_network_arch} is trained using Stochastic Gradient Descent (SGD) optimizer with a learning rate of 0.001 for 15000 epochs. Knowledge distillation is also used during the elastic training stages as described in the original OFA paper \cite{cai2020once}. It fuses two loss terms using both the soft labels of the supernet just before initialization of the elastic training stages, and the real labels \cite{hinton2015distilling}. In Fig. \ref{fig:nas_training_plot}, the progress of the training is given in terms of the objective loss with respect to both training epochs and time spent. The elastic kernel, elastic depth, and elastic width stages are also given in this graph. An evolutionary search algorithm is run using this model to find the best-performed model given in Fig. \ref{fig:nas_training_plot}. Note that the details of the search algorithm and selected parameter values are given in our project repository \cite{ai8xtraining_repo}.

The selected model parameters must be quantized to at most 8-bits integers in order to be deployed to our edge platform described in the following section. However, post-quantization of the trained networks causes high-performance degradation, and it is much higher when the model is deeper or includes batch normalization, according to our experiments. Therefore, we use the second step of training, Quantization Aware Training (QAT) of the selected model, to fine-tune the model parameters while minimizing the induced errors due to the quantization. The QAT approach implements the fake (simulated) quantization approach in \cite{Jacob_2018_CVPR}. In this training step, an Adam optimizer is used with an initial learning rate of 0.001, gradually halved 5 times starting from the 100th epoch. The model is trained for 200 epochs, and the QAT procedure is initiated after epoch 150. In this study, all model parameters are quantized to 8-bits, and existing batch normalization layers are folded into the weights with the initialization of QAT.

\subsection{Neural Network Accelerator}
As a deployment platform of the proposed CNN model, MAX78000 is chosen. MAX78000 is an
Ultra-Low-Power Arm Cortex-M4 processor with FPU-based MCU with CNN accelerator. By spending only microjoules of energy to execute inferences, MAX78000 targets battery-powered neural network-based applications at the edge \cite{max78000_datasheet}. The CNN engine supports both Conv1D and Conv2D as convolution operations, 2D transposed convolution layer, linear layer, pooling layers, and element-wise operations as addition, subtraction, binary OR, binary XOR.

% The MAX78000 [1] is a new breed of Artificial Intelligence (AI) MCU built to enable neural networks to execute at ultra-low power and live at the edge of the IoT. This product combines the most energy-efficient AI processing with Maxim's proven ultra-low-power MCUs. The hardware-based CNN accelerator enables battery-powered applications to execute AI inferences while spending only microjoules of energy. This makes it an ideal architecture for keyword spotting applications. The MAX78000 features an Arm® Cortex®-M4 with FPU CPU for efficient system control with an ultra-low-power deep neural network accelerator.

\subsection{Accuracy Results}
Table \ref{tab:accuracy_comparison} summarizes the results of our findings of the proposed KWS approach with the selected small footprint models in the literature. We use class accuracy as the evaluation metric to be consistent with the literature. However, these studies consider the footprint of the approach by only taking the model complexity into account without considering the feature extraction step, i.e., MFCC calculation. Therefore, we present the table with a field that shows whether the approach uses MFCC or not. The number of parameters in the model, as well as the precision of those parameters and activations, are given in the table.

\begin{table}
    \caption{Comparison of accuracy results on the Speech Command Dataset}
    \begin{center}
 \begin{tabular}{c c c c c c} 
 \textbf{Model} & \textbf{MFCC} & \textbf{Precision} & \textbf{\#Param} &
 \textbf{Model Size} &
 \textbf{Acc} \\ [0.5ex] 
 \hline
 \cite{sainath2015convolutional} & yes & fp-32 & 244.2K & 976.8 KB & 90.2\% \\ 
 
 \cite{tang2018deepresidual} & yes & fp-32 & 238K & 952 KB & 95.8\% \\
 
 \cite{chen2019smallfootprint} & yes & fp-32 & 61K & 244 KB & 96.4\% \\
 
 \cite{zhang2018hello} & yes & int-8 & 497.6K & 497.6 KB & 95.4\% \\
 
 \cite{mittermaier2020kwswithsincconv} & no & fp-32 & 122K & 488 KB & 96.6\% \\
 
 ours & no & \textbf{int-8} & 419.8K & \textbf{419.8 KB} & \textbf{96.3\%} \\ [1ex] 
 
\end{tabular}
\end{center}
    \label{tab:accuracy_comparison}
\end{table}

The proposed model in this study provides an accuracy of over 96\%, which is above or comparable with the state-of-the-art small footprint KWS models despite not utilizing pre-calculated MFCCs. Having similar accuracies with the state-of-the-art proves that specialized hardware for Neural Net inference can easily be used to run the pre-trained models for KWS applications on the edge as the model runs on the raw audio signal.

\subsection{Energy Measurements}
In order to analyze the efficiency provided by MAX78000 and our approach, two sets of measurements are presented in this study. The first measurement is made to calculate the latency and the energy consumption of the MAX78000 for the proposed model. The same model is also implemented for an ultra-low-power ARM Cortex M4F microprocessor, and the measurements are given in \ref{tab:energy_comparison}. It should be noted that the same model consumes more than 40 times more energy when deployed in the ARM Cortex M4F processor compared to MAX78000 CNN accelerator. One major element in this low consumption figure is that our model does not need any preprocessing outside the accelerator; therefore, it can better leverage the accelerator's energy efficiency.

\begin{table}
    \caption{Comparison of latency and energy measurements of the proposed model for different hardware}
    \begin{center}
 \begin{tabular}{c c c c c} 
 \textbf{Device} & \textbf{Latency (ms/inf)} & \textbf{Energy (uJ/inf)} \\ [0.5ex] 
 \hline
 MAX78000 & 3.5 & 251 \\ 
 
 ARM Cortex M4F & 905 & 11200\\ [1ex] 
 
\end{tabular}
\end{center}
    \label{tab:energy_comparison}
\end{table}

Secondly, we measure the energy and latency for MFCC calculation to show the efficiency of the proposed approach for the state-of-the-art approaches as they use MFCC transformation as an initial feature extraction procedure. 40 MFCC features extracted from a speech frame of length 40ms with a stride of 20ms, ending with a 49x40 data for a 1s long audio. Note that this is the same configuration used in \cite{zhang2018hello} and similar to the others used in the literature. The described MFCC feature extraction is measured to consume 826 uJ, more than three times what our end-to-end KWS model consumes in the MAX78000 accelerator. This means any conventional model in the literature that is built upon MFCC features consumes much more energy than our solution, even if it is deployed in the convolution efficient hardware.

% \textcolor{red} {
% Measurments:
% \begin{itemize}
% \item Our model on MAX78000: 251 uJ (Data Transfer 11uj + Inference 240 uJ)
% \item Our model on ARM Cortex M4F: 11.4 mJ
% \item MFCC on ARM Cortex M4F: 826 uJ
% \end{itemize}
% }

% \textcolor{blue} {This is the only missing part in the paper. We will complete this part once we performed the energy calculations.}

% \textcolor{blue} {
% TODOs:
% }
% \begin{itemize}
% \color{blue}
%   \item Energy & latency comparison KWS on MAX78000 vs. ARM M4F with CMSIS-NN
%   \item Energy & latency comparison for MFCC calculation vs. our model evalaution
% \end{itemize}
% \textcolor{blue} {
% Here is an example for how to show the listed energy requirement comparisons:
% }
% \begin{figure}[h]
% \centering
% \includegraphics[trim=0 0 0 0,clip, width=1.01\linewidth]{figures/energy_example.PNG}
% \caption{Sample energy measurment graph}
% \label{fig:energy_example}
% \end{figure}

\section{Conclusion}
In this paper, we present a highly accurate and energy-efficient KWS solution that runs on battery-powered devices. For this purpose, we propose a novel CNN model that works on raw audio by eliminating commonly used MFCC preprocessing. By doing so, the whole of the proposed network becomes deployable to the NN accelerators. In order to find the best architecture, we adopted a weight-sharing-based NAS approach, OFA, and we show that our accuracy is comparable with the state-of-the-art small footprint KWS solutions. Lastly, we deploy the model to a NN accelerator, MAX78000, and present its effectiveness in terms of energy and latency.

\section{Acknowledgements}
The authors would like to thank Robert Muchsel, Brian Rush and other members from the AI Development Group at Analog Devices that
contributed to this work.

% \section*{References}
\bibliographystyle{./bibliography/IEEEtran}
\bibliography{./bibliography/references}

\end{document}